\documentclass{ws-procs9x6}

\begin{document}

\title{IceCube: The Cubic Kilometer Neutrino Telescope at the South Pole}

\author{A.~R. FAZELY for the IceCube Collaboration\footnote{\uppercase{F}ull 
author list is given at the end of the paper}}

\address{Department of Physics, \\
Southern University and A\&M College, \\ 
Baton Rouge, Louisiana 70813\\ 
E-mail: fazely@phys.subr.edu}

\maketitle

\abstracts{Search for ultra high-energy neutrino induced reactions, as part 
of a comprehensive probe of the neutrino sky and also investigation of 
the particle nature of the dark matter, with unique sensitivity to cold dark 
matter particles are described. We present a description of the
design, scientific motivation and goals, performance and status of the IceCube
experiment.}     

\section{Introduction}

The main motivation for the IceCube experiment \cite{IceCube} is to
probe the universe with ultra high energy (UHE) neutrinos and to search for the
signature of cold dark matter. The IceCube 
detector 
will provide astrophysical and particle physics information, essential
to the understanding of the origin of the highest energy cosmic rays as well as a test of the fundamental laws of physics. 

\begin{figure}[ht]
\epsfxsize=10cm   
\centerline{\epsfxsize=4.1in\epsfbox{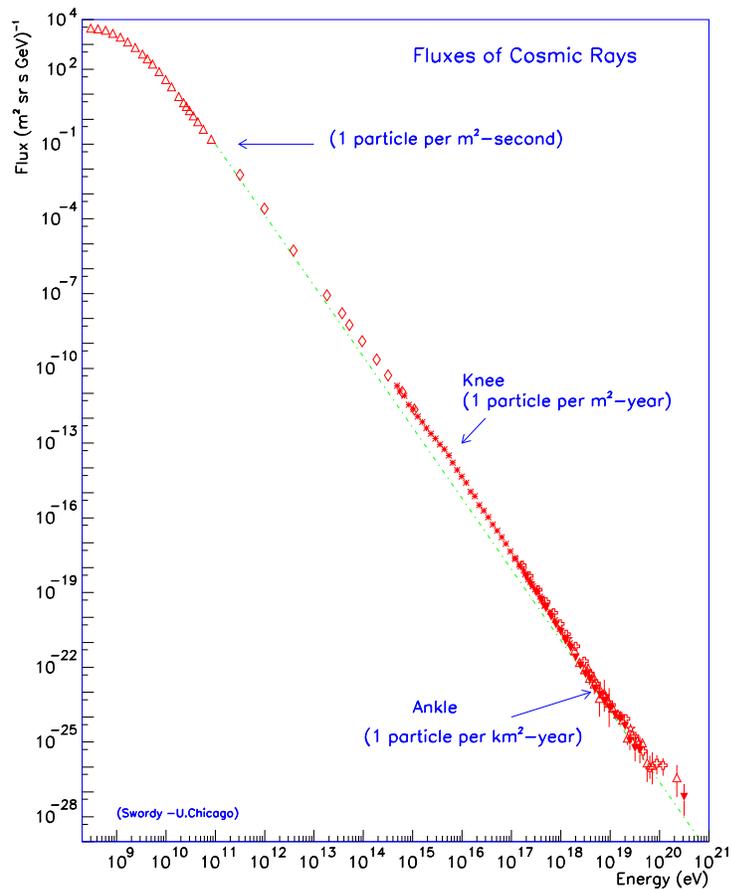}}
\caption{All-particle energy spectrum
\label{sciam1}}
\end{figure}

The all-particle spectrum, as shown in Figure~\ref{sciam1}, is dominated by two
main features at 3 PeV and at 5 EeV commonly referred to as the "knee" and the
"ankle". The spectrum shows a steep drop in the
flux of cosmic rays as a function of energy. The slope becomes steeper at the 
knee and rises at the ankle. Many attempts have been made to explain the drop 
and increase in the flux at the knee and the ankle, respectively with 
diverse fortune (see for example Esteban Roulet \cite{lpc} 
and references therein).  

In the UHE region of the knee,
the particle flux drops to $1/m^2-year$ and falls to $1/km^2-year$ at the
ankle. These fluxes are impossible to observe by conventional
detectors. The detection of these particles with reasonable statistics will
provide the necessary
information regarding sources and the nature of these particles. Therefore,
at the scale of the IceCube detector, one can begin to perform 
efficient neutrino detection in the
PeV region, above where the "knee" in the all-particle spectrum occurs.

IceCube is designed to search for sources of UHE neutrinos such as Active 
Galactic Nuclei (AGN), Supernova Remnants (SNR) or micro-quasars, and
neutrinos from Gamma Ray Bursts (GRB). The sensitivity of
IceCube to astrophysical sources has been studied in reference
\cite{IceCubeapp}.

Aside from a search for Astrophysical sources of neutrinos,
IceCube can also provide answers to a series of questions, related to particle 
physics,
such as search for neutrinos from possible
candidates for cold dark matter, weakly interacting massive particles (WIMPs)
annihilating in the sun, and magnetic monopoles or other exotic particles
such as strange quark matter or Q-balls predicted by SUSY models
\cite{Halzen}. Furthermore, IceCube with its cubic kilometer size is
able to examine a possible enhancement in neutrino interaction cross 
sections due
to extra dimensions where graviton contributions could
increase the total neutrino cross section to the level of hadronic interaction
cross
sections, of the order of tens of mb\cite{muniz}.   

\section{IceCube Detector Setup}

\begin{figure}[ht]
\begin{center}
\epsfxsize=10cm  
\includegraphics[width=.55\textwidth,angle=-90]{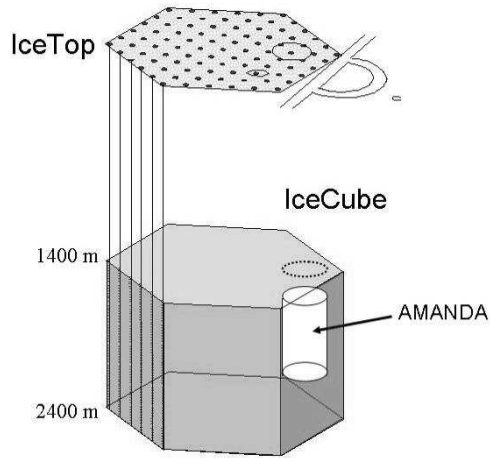}
\caption{IceCube, IceTop and AMANDA} 
\label{IceCube2}
\end{center}
\end{figure}

The IceCube detector layout is shown in Figure~\ref{IceCube2}. IceCube consists
of 4800 optical modules (OM) mounted on 80 strings regularly spaced such that 
each two adjacent
strings are 125 m apart.
In planar view, IceCube covers an area of approximately 1 km$^2$.
The instrumented part of the string is at a depth of 1,450 to 2,450 m below
the surface of the ice. Each PMT string
consists of 60 OM's, with OM's equally spaced at a distance of 
approximately 17 m.
The strings are
arranged in a hexagonal pattern in planar view.
At the ice surface, on the top of each string, a
station of the IceTop air shower array will be installed.
An IceTop station consists of two ice tanks with a total area of 
7 m$^2$. Two of these tanks have been installed during the
2003-04 season at the South Pole. The IceTop will be operated in coincidence
with the in-ice arrays. This provides a veto for the downward going events as 
well as information on the chemical composition of the cosmic rays up to
$10^{18}$eV \cite{spier}. 
The AMANDA-II detector \cite{amanda}
will be integrated into IceCube. The present configuration is designed for 
optimum sensitivity to muon neutrinos in the energy range of 1-100 TeV.\\

\section{The Digital Optical Module}

\begin{figure}[ht]
\begin{center}
\includegraphics[width=.55\textwidth]{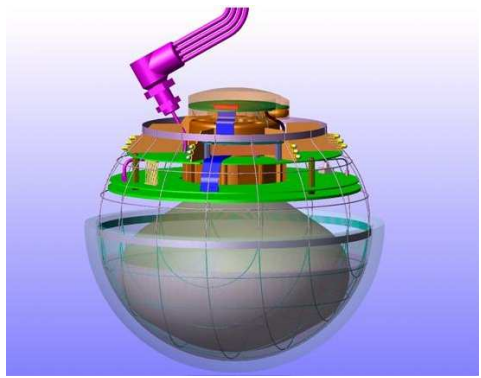}
\caption{
Schematic view of the IceCube DOM}
\label{DOM}
\end{center}
\end{figure}

The heart of the Data Acquisition system of IceCube is the Digital
Optical Module (DOM). The IceCube DOM shown in Figure~\ref{DOM} 
contains a 10-inch diameter PMT, HAMAMATSU R-7081. These PMT's provide 
excellent charge and time
resolution. The dynamic range is 200 photo-electrons (pe)/10 nsec, with an integrated dynamic range of more than 2000 pe's. The signal is digitized at the PMT 
level with a digitization depth of 4 $\mu$-sec. A single PMT low noise 
rate of less than  
500 Hz, at operating temperatures of the South Pole 
of -20 to -40 $^o$C has been achieved with these PMT's. Each PMT and its 
associated electronics is housed in a glass sphere that can withstand a pressure
of 10,000 psi (68,948 kPa). The face of the PMT makes contact with this glass 
shell through a gel with approximately the same index of refraction as that of 
the glass. 

\section{IceCube Performance}

\begin{figure}[ht]
\begin{center}
\includegraphics[width=.65\textwidth]{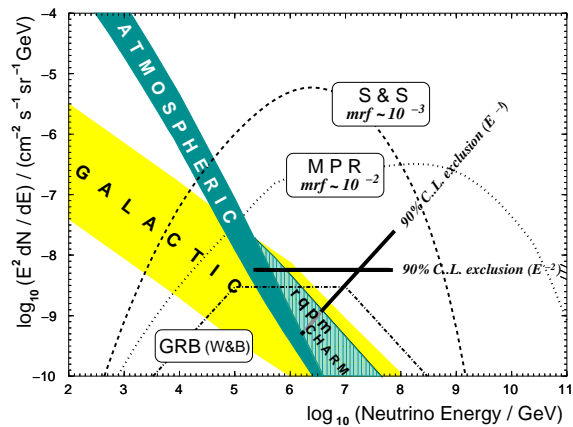}
\caption{
Expected sensitivities of the IceCube detector. See text for
explanations.}
\label{diffuse}
\end{center}
\end{figure}

The main particle detection capability of the IceCube detector is measured in 
terms of its sensitivity to the detection of muons. Muons are produced in the
detector because of charged-current interactions of $\nu_\mu$ with H and O 
nuclei in the ice or other nuclei in the ground below the ice. The majority 
of the downward-going muons in IceCube are due to the interaction of the 
primary cosmic rays with the atmosphere.     
Figure~\ref{diffuse} summarizes
the expected sensitivity to diffuse fluxes as    
a function of neutrino energy.
Solid lines show the expected 90\% confidence level (CL) limits for
$E^{-2}$ and $E^{-1}$ fluxes, respectively. These calculations assume 
a period of three years of data accumulation.  
The dashed lines show the model proposed by Stecker and Salamon describing the
photo-hadronic interactions in the AGN core\cite{SS}. The dotted
line shows the model of Mannheim, Protheroe, and Rachen, 
estimating neutrino emission from photo-hadronic interactions in AGN
jets \cite{MPR}.  Also shown is the GRB
estimate of Waxman and Bahcall \cite{WB}.
Their estimate yields approximately ten GRB events for 1000 GRB's monitored.   
IceCube performance at higher
energies is described by Yoshida, Ishibashi, and Miyamoto. They show that in the
10 - 100 PeV energy range, not only muons, but also $\tau$'s survive without 
decay and would leave detectable signals in horizontal and downward-going events\cite{Yoshida}.

\vspace{-1mm}
\section{IceCube Deployment}

The hot-water drill technology has been developed and perfected over decades as
the fastest and the most efficient method for drilling holes in the ice. 
It turns
out that for IceCube hot water also provides the only possible technology. 
Water is 
readily available and the water in the hole after deployment of the strings
freezes, preserving the optical properties of the ice. The optics for IceCube 
are well understood from years of experience with AMANDA.
Holes with 60 cm
diameter will be drilled with 100 $^o$C water at a rate of 16 holes per year. 
The drill system is an evolution of the AMANDA drill called the
Enhanced Hot Water Drill (EHWD).
The EHWD drill is energy intensive because of the large amount of energy required for the ice-to-water phase transition. The total time required to drill a 
2400 m deep hole is about 40 hours. The power consumption for  
the EHWD will be 5 MW, compared to 2 MW for AMANDA.
This, and the larger diameter and the length of the water transporting
hoses, will result in 40 hours needed to drill a
2400 m deep hole.
Mounting, testing and deployment of a string with 60 DOM's
are estimated to take about 20 hours. 
                                      
\section{IceCube Status}
The IceCube collaboration includes about 150 scientists from
institutions in Belgium, Germany, Japan, Netherlands, New Zealand,
Sweden, UK, US and Venezuela. In the US IceCube is funded by the
National Science Foundation through a multi-year Major Research Equipment 
(MRE) grant. The start-up funding for a period of two years from the NSF 
began in 2002. This year (2004) IceCube has begun its implementation phase as a 
fully funded NSF, MRE. Furthermore, significant funding for the IceCube project 
has been approved in Belgium, Germany and Sweden.
Full IceCube construction has begun this year and will take 6 years. The first 
parts of the EHWD's have been shipped to the pole. This year the production and 
testing of 150 DOM's have been scheduled.  
The drilling of the first holes is scheduled to begin in January
2005. 
\section*{Acknowledgments}
This work was supported by the National Science Foundation through a subcontract from the University of Wisconsin, Madison.

%
%
%
%

\section*{IceCube Author List}

\vspace{-1mm}
\begin{sloppypar}
\noindent
{\scriptsize
A.~Achterberg$^{26}$, M.~Ackermann$^{6}$ J.~Ahrens$^{15}$, 
J.N.~Bahcall$^{8}$,
X.~Bai$^{1}$, R.C.~Bay$^{14}$, T.~Becka$^{15}$, 
K.-H.~Becker$^{2}$,
J.~Bergmans$^{26}$, D.~Berley$^{16}$
% ...
}    

\end{sloppypar}
{\footnotesize
\noindent
   (1) Bartol Research Institute, University of Delaware, Newark, DE 19716, USA
   \newline
   (2) Fachbereich 8 Physik, BUGH Wuppertal, D-42097 Wuppertal, Germany
   \newline
   (3) Universit\'e Libre de Bruxelles, Science Faculty CP230, Boulevard du Trio
mphe, B-1050 Brussels, Belgium
   \newline
   (4) CTSPS, Clark-Atlanta University, Atlanta, GA 30314, USA
   \newline
   (5) Dept. of Physics, Chiba University, Chiba 263-8522 Japan
   \newline
   (6) DESY-Zeuthen, D-15738 Zeuthen, Germany
   \newline
   (7) Blackett Laboratory, Imperial College, London SW7 2BW, UK
   \newline
   (8) Institute for Advanced Study, Princeton, NJ 08540, USA
   \newline                                                             
  (9) Dept. of Physics and Astronomy, University of Kansas, Lawrence, KS 66045,
 USA
   \newline
   (10) Lawrence Berkeley National Laboratory, Berkeley, CA 94720, USA
   \newline
   (11) Dept. of Astronomy and Astrophysics, Pennsylvania State University, Univ
ersity Park, PA 16802, USA
   \newline
   (12) Dept. of Physics, Pennsylvania State University, University Park, PA 168
02, USA
   \newline
   (13) Dept. of Physics, Southern University, Baton Rouge, LA 70813, USA
   \newline
   (14) Dept. of Physics, University of California, Berkeley, CA 94720, USA
   \newline
   (15) Institute of Physics, University of Mainz, Staudinger Weg 7, D-55099 Mai
nz, Germany
   \newline
   (16) Dept. of Physics, University of Maryland, College Park, MD 20742, USA
   \newline
   (17) Dept. of Physics, George Mason University, Fairfax, VA 22030, USA
   \newline
   (18) University of Mons-Hainaut, 7000 Mons, Belgium                         
   \newline
   (19) Department of Physics, University of Oxford, Oxford, England 
   \newline
   (20) Departamento de F\'{\i}sica, Universidad Sim\'on Bol\'{\i}var, Caracas,
1080, Venezuela
   \newline
   (21) Dept. of Astronomy, University of Wisconsin, Madison, WI 53706, USA
   \newline
   (22) Dept. of Physics, University of Wisconsin, Madison, WI 53706, USA
   \newline
   (23) SSEC, University of Wisconsin, Madison, WI 53706, USA
   \newline
   (24) Physics Dept., University of Wisconsin, River Falls, WI 54022, USA
   \newline
   (25) Division of High Energy Physics, Uppsala University, S-75121 Uppsala, Sweden
   \newline
   (26) Faculty of Physics and Astronomy, Utrecht University, NL-3584 CC Utrecht
, The Netherlands
   \newline
   (27) Dept. of Physics, Stockholm University, SE-10691 Stockholm, Sweden
   \newline
   (28) Vrije Universiteit Brussel, Dienst ELEM, B-1050 Brussels, Belgium
   \newline                      
}


\end{document}